\documentclass[prb,final,showpacs,twocolumn,superscriptaddress,aps]{revtex4-2}

\usepackage[utf8]{inputenc}
\usepackage[american,british]{babel}
\usepackage[T1]{fontenc}
\usepackage[pdftex]{graphicx}  
\usepackage{graphicx, xcolor}
\usepackage{dcolumn}
\usepackage{braket}
\usepackage{bm}
\usepackage{amsmath,amsthm,amssymb}
\usepackage{color}
\usepackage{verbatim}
\usepackage{lipsum}

\usepackage[colorlinks,linkcolor=blue,citecolor=blue,urlcolor=blue]{hyperref}

\usepackage{dsfont}
\usepackage{physics}
\usepackage{soul}

\newcommand{\commentas}[1]{{\textcolor{black}{#1}}}

\def \be {\begin{equation}} 
\def \ee {\end{equation}} 
\def \l {\left(} 
\def \r {\right)} 
\def \la {\langle} 
\def \ra {\rangle}  
\def\lc#1{{ \color{black}  #1}}

\begin{document}
\title{Spreading of a local excitation in a Quantum Hierarchical Model}

\author{Luca Capizzi}
\email{lcapizzi@sissa.it}
\affiliation{SISSA, via Bonomea 265, 34136 Trieste, Italy}
\affiliation{INFN Sezione di Trieste, via Bonomea 265, 34136 Trieste, Italy}

\author{Guido Giachetti}
\email{ggiachet@sissa.it}
\affiliation{SISSA, via Bonomea 265, 34136 Trieste, Italy}
\affiliation{INFN Sezione di Trieste, via Bonomea 265, 34136 Trieste, Italy}

\author{Alessandro Santini}
\email{asantini@sissa.it}
\affiliation{SISSA, via Bonomea 265, 34136 Trieste, Italy}

\author{Mario Collura}
\email{mcollura@sissa.it}
\affiliation{SISSA, via Bonomea 265, 34136 Trieste, Italy}
\affiliation{INFN Sezione di Trieste, via Bonomea 265, 34136 Trieste, Italy}

\date{\today}

\begin{abstract}
We study the dynamics of the quantum Dyson hierarchical model in its paramagnetic phase. An initial state made by a local excitation of the paramagnetic ground state is considered. We provide analytical predictions for its time evolution, solving the single-particle dynamics on a hierarchical network. A localization mechanism is found and the excitation remains close to its initial position at arbitrary times. Furthermore, a universal scaling among space and time is found related to the algebraic decay of the interactions as $r^{-1-\sigma}$.
We compare our predictions to numerics, employing tensor network techniques, for large magnetic fields, discussing the robustness of the mechanism in the full many-body dynamics. 

\end{abstract}

 
\maketitle

\section{Introduction}

Long-range-interacting systems, characterized by slow-decaying power-law potentials, are known to exhibit a plethora of peculiar behaviors \cite{campa2009physics}.
Among all, dynamical phase transitions\,\cite{zhang2017observation, baumann2010dicke}, long-lived metastable states\,\cite{campa2009physics,defenu2021metastability,Santini_2022}, time crystalline phases\,\cite{rovny2018observation, choi2017observation, zhang2017observation,ArxivDTCMario,Giachettitimecrystals2022}, peculiar critical properties in low dimensions \cite{giachetti2021bkt, defenu2015fixed}, exotic defect scaling\,\cite{safavi2018verification,keesling2019quantum}, slow entanglement propagation \cite{lanyon2013entanglement,pappalardi2018scrambling,lerose2020origin, giachetti2021entanglement,scopa2021entanglement}. These phenomenology stimulated an impressive theoretical activity aimed at understanding the equilibrium and non-equilibrium behavior of such systems\,\cite{monroe2019programmable,mivehvar2021cavity,defenu2021longrange}. Examples of long-range physics can be observed across several fields of physics from astrophysics\,\cite{lyndenbell1967statistical,campa2009physics,giachetti2020coarse}, to plasma physics\,\cite{barre2012algebraic} and fluid dynamics \cite{onsager1949statistical}; they also have been engineered in experimental setups based on molecular and optical systems (AMO)\,\cite{defenu2021longrange}, trapped ions\,\cite{monroe2019programmable, britton2012engineered}, Rydberg gases\,\cite{baranov2012condensed} and optical cavities\,\cite{landig2016quantum,mivehvar2021cavity}. 

In this context, the classical hierarchical model was originally introduced by Dyson in his seminal paper \cite{dyson1969existence} as a tool for understanding the critical properties of one-dimensional long-range spin systems. Here the usual translational-invariant adjacency matrix of the couplings is replaced by those of a hierarchical network, allowing to explicitly carry out the renormalization group procedure \cite{Kim1977critical}. A quantum counterpart of the hierarchical Ising model, in which the classic spin variables are replaced with non-commuting spin operators, has been proposed in \cite{Monthus_2011}. The strong disorder renormalization group (SDRG) \cite{igloi2018strong} can be used to understand the ground-state properties of the hierarchical quantum Ising model in a transverse magnetic field \cite{Monthus_2016}, and its entanglement content \cite{Pappalardi_2019}.

We also mention that a relation between the hierarchical models and a field theoretical formulation in fractal spaces has been proposed, and it is still a current topic of research in the context of high-energy physics. In particular, we refer to the characterization of adelic string amplitudes \cite{Freund:1987ck} together with a $p$-adic formulation of the AdS/CFT principle \cite{Gubser:2016guj,Heydeman:2016ldy,
Hung:2019zsk}, where the complex field $\mathbb{C}$ is replaced by the $p$-adic field $\mathbb{Q}_p$.

In this work, we focus on the 
dynamics of the 
hierarchical quantum Ising model in the paramagnetic regime. We consider an initial state made by a localized excitation and we study its time evolution. We find that the lack of translational invariance, replaced here by a hierarchical tree structure, results in the localization of the excitation around its initial position. In addition, thanks to the presence of a self-similar structure of the hamiltonian, a scaling relation between space and time ($t\sim x^z$) is found, 
which differs from the one at criticality~\cite{Monthus_2011}.

The manuscript is organized as follows: in Sec.~\ref{sec:Model} we introduce the 
model as a quantum chain with tree-structured long-range interactions.
In Sec.~\ref{sec:Loc} we focus on the dynamics of the localized excitation in the paramagnetic phase, which turns out to be equivalent to a Schroedinger equation in the presence of a hierarchical long-range hopping. The evolution of the single-particle wave function is thus constructed explicitly and its properties of localization are discussed rigorously. 
In Sec.~\ref{sec:Num} we compare our analytical predictions, which are exact in a proper scaling limit, with the evolution of a finite chain, obtained via  numerical techniques, and we found a very good agreement.
Finally, we draw our conclusions in Sec.~\ref{sec:Con} and report some additional details in the appendices.

\section{Model}\label{sec:Model}
\begin{figure}[t]
    \centering
    \includegraphics[width=1\linewidth]{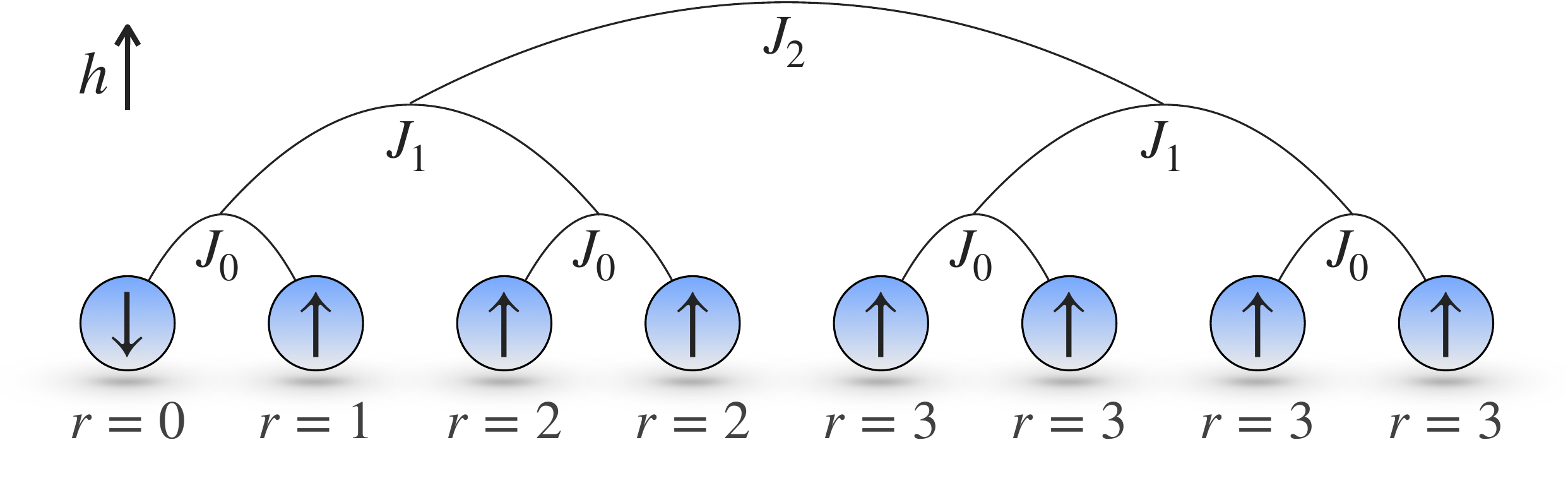}
    \caption{Schematic representation of the hierarchical Dyson model for $N=3$, where the branches of the binary tree highlight the structure of the interactions. We show the hierarchical distances from the first site $r\equiv r(1, j)$.}
    \label{fig:HDM_scheme}
\end{figure}
We consider a one-dimensional spin-$1/2$ lattice of length $L=2^N$. We can arrange the spins in $N+1$ possible binary partitions $\pi_p$ whose elements are composed of collections of consecutive and adjacent $2^p$ spins. We may define with $(j, ...,j')$ an element of a certain partition $\pi_p$ which contains all the spins from the position $j$ to $j'$. In particular, at the lowest level $p=0$, the partition $\pi_0$ contains $L$ elements $(1), (2), ..., (L)$. Then, $\pi_1$ contains $L/2$ blocks $(1,2),(3,4), ...,(L-1,L)$, again, $\pi_2$ contains $L/4$ blocks $(1,2,3,4),(5,6,7,8), ...., (L-3,L-2,L-1,L)$, and so on, up to the final partition $\pi_N$ whose only element is the whole chain $(1,2,3, ..., L)$. 

We can distinguish an element of a partition by a pair $(p, q)$, $p$ identifies the partitioning level while $q$, $q=1, ...,2^{N-p}$, runs over the elements of $\pi_p$. For each element $(p,q)$ 
we identify the collective spin 
\begin{equation}
    \mathbf{S}_{(p,q)} \equiv \sum_{j} \boldsymbol{\sigma}_j,\quad (q-1)2^p +1 \leq j\leq q2^p,
\end{equation}
where $\boldsymbol{\sigma}_j = (\sigma^x_j,\sigma^y_j,\sigma^z_j)$ is the vector of the Pauli matrices at position $j \in \{1,\dots,L\}$. 

We can now define the Hamiltonian of the Quantum Hierarchical Dyson Model (HDM) as\begin{equation}
    H = - \sum_{p=0}^{N-1}\;\sum_{q=1}^{2^{N-p-1}} J_p \, S^x_{(p,2q-1)}S^x_{(p,2q)} - h S^z_{(N,1)},\label{eq:hierarchical_hamiltonian}
\end{equation}
where
$
J_p = J/2^{(1+\sigma)p}
$
represents the interaction term at level $p$. A sketch of the model is schematically represented in Fig. \ref{fig:HDM_scheme}. 
Let us notice that the term proportional to $h$ represents the coupling with a uniform transverse magnetic field since $ S^{z}_{(N,1)} = \sum_{j=1}^{2^{N}} \sigma^{z}_{j}$,
while the longitudinal interaction along the $x$-axis among distinct spins has been introduced so that it displays a hierarchical structure. For the sake of convenience, we define a hierarchical distance
$
r(i,j)
$
as the minimum level $p$ for which the sites $i$ and $j$ belong to the same element of the partition $\pi_p$ (e.g. $r(1,1)=0, r(1,2)=1, r(1,3)=r(1,4) =2$ and so on). In this way, one can rewrite the Hamiltonian~\eqref{eq:hierarchical_hamiltonian} as\begin{equation}
    H = -\sum_{i<j} J_{r(i,j)-1} \, \sigma^x_i\sigma^x_{j} - h \sum_{i} \sigma_i^z,\label{eq:hierarchical_hamiltonian1}
\end{equation}
explicitly showing how the interaction among different spins depends on the distance $r(i,j)$, instead of the euclidean one $\abs{i-j}$. 
To relate these two quantities, we may roughly estimate $\abs{i-j}\approx 2^{r(i,j)}$, which means that the coupling strength of the model scales as
\begin{equation}
    J_{r(i,j)-1}=\frac{J}{2^{(1+\sigma)(r(i,j)-1)}} \approx \frac{J}{\abs{i-j}^{1+\sigma}}.
\end{equation}
From now on we will restrict the analysis of the model to $\sigma > 0$.

\section{Localization}\label{sec:Loc}

In this section, we investigate the dynamics of the model in the paramagnetic phase, and in particular in the limit of a large magnetic field
$
h \gg J
$.
Due to the separation of scales, states with distinct values of the total magnetization along $z$ 
are effectively decoupled, 
and thus we may refer to $S^z_{(N,1)}$ as a quasi-conserved charge.
Above the fully polarised paramagnetic state, namely
$
\ket{\uparrow \dots \uparrow}
$,
one can create excitations via local spin flips. 
We interpret these excitations as states with particles localized at some points along the chain. In this regime, the dynamics is equivalent to the one 
induced by the following hard-core bosons (effective) Hamiltonian
\begin{equation}
     H = -\sum_{i<j} J_{r(i,j)-1} \left(b^\dagger_ib_j+b^\dagger_jb_i\right),\label{eq:hard_core}
\end{equation} 
where $b_i^\dagger,b_i$ are the creation/annihilation operator of hardcore bosons, satisfying
$
[b_i,b_j] =[b^\dagger_i,b^\dagger_j]=0 
$,
$
[b_i,b^\dagger_j]=\delta_{i,j}
$,
with the additional constraint $b^2_j=(b_j^\dagger)^2 = 0$.
Notice that, even if this description is exact only for $h/J \to +\infty$, the qualitative picture is kept unchanged as long as $h/J$ is large enough to ensure that the system is sufficiently deep in the paramagnetic phase (see \cite{Monthus_2015} for further details).

We are interested in the dynamics induced by a strongly paramagnetic Hamiltonian after a local excitation (spin-flip) has been created on top of the fully polarised state. 
Namely, we prepare the system in the initial configuration 
$
\ket{\psi(0)} = \ket{\downarrow \uparrow\uparrow\dots \uparrow}
$,
and then we let it evolve in time, thus analyzing the spreading of a single particle initially localized in the first lattice site.  
Let us mention that the position of the initial spin flip is unimportant since the chain is homogenous: indeed, despite the lack of translational symmetry, the Hamiltonian is invariant under permutations of sites that keep fixed the hierarchical distance $r(i,j)$.

\lc{The protocol under consideration is simple enough to be exactly solvable, and the thermodynamic limit can be analyzed. We expect that the resulting picture is able to capture the salient features of the whole paramagnetic phase, and it should be predictive if the particles are diluted enough. Moreover, some crucial properties, which depend mostly on the tree-structure of the Hamiltonian, are highlighted and possibly they are shared by any hierarchical model, irrespective of the microscopic details.} In this regime, the dynamics remains in the single-particle sector,
and the projected Hamiltonian (\ref{eq:hard_core}) can be exactly diagonalised.

In the following, to simplify the notation,  we introduce $r\equiv r(1,j)$,
as the hierarchical distance between the $j\rm$th and the first site. Moreover, we describe a generic one-particle state $\ket{\Psi(t)} = \sum_{x=1}^{L} \psi(x,t) \ket{x}$ via its associated wave-function
$\psi(x,t) = \braket{x}{\Psi(t)}$, where $\ket{x}$ represents a state with a single particle (spin-flip) localized at position $x$ on the chain. As a consequence, the initial state $\ket{\Psi(0)}$ is simply characterised by the delta-peaked wave-function $\psi(x,0) = \delta_{x,1}$.
We can decompose further the inital wave-function in terms of the eigenfunctions of single-particle Hamiltonian
(see Appendix~\ref{app:Diagon}), obtaining
\begin{align}
\psi(x,0) = &\frac{1}{L}\chi_{[1,L]}(x)+\\&\frac{1}{L}\sum^{N}_{k=1} 2^{k-1} \l \chi_{[1,L/2^k]}(x) -\chi_{[1+L/2^k,L/2^{k-1}]}(x) \r,
\label{eq:delta_expansion}
\end{align}
where $\chi_{[a,b]}(x)$ is the characteristic function of the interval $[a,b]$ defined as
\be
\chi_{[a,b]}(x) = \begin{cases} 1 \quad x \in [a,b],\\ \ \\0 \quad  \text{otherwise}.
\end{cases}
\ee
Through the previous decomposition, we can easily express the time-evolved state as
\begin{align}\label{eq:1part_wavef}
&\psi(x,t) =\frac{1}{L}\chi_{[1,L]}(x)e^{-i\epsilon_0t}+\\ &\frac{1}{L}\sum^{N}_{k=1}e^{-i\epsilon_kt}2^{k-1} \l \chi_{[1,L/2^k]}(x) -\chi_{[1+L/2^k,L/2^{k-1}]}(x) \r,\notag
\end{align}
with $\epsilon_k$ being the single-particle energies 
\begin{align}\label{eq:eps_eig}
\epsilon_k = -\frac{J}{1-2^{-\sigma}}\l 1 - \frac{2^{k\sigma}}{L^\sigma}\r + J \ \frac{2^{k\sigma}}{L^\sigma}\left(1-\delta_{k,0}\right).
\end{align}
Here $k=0,\dots, N$ labels, in ascending order, only the distinct eigenvalues of the coupling matrix, whose dimension is $2^N$. Apart from the first two non-degenerate eigenvalues, it can be shown (see Appendix~\ref{app:Diagon}) that all the others have degeneracy $2^{k-1}$, such that $1+\sum_{k=1}^{N}2^{k-1} = 2^{N}$.

Since the value of the wave-function at position $x$ depends only on the hierarchical distance $r =  \lfloor \log_{2}(x) \rfloor $, it is convenient to write it explicitly as a function of $r$ and $t$ only, denoting it as $\psi(r,t)$ (with a slight abuse of notation). For $r>0$ its expression reads as follows 
\begin{align}
 \psi(r,t) = \frac{1}{L} e^{-i \epsilon_0 t} +\frac{1}{L} \sum^{N-r}_{k=1} 2^{k-1} \ e^{- i \epsilon_kt} - 2^{-r} e^{- i \epsilon_{N-r+1} t},
 \label{eq:psi_generic_site} 
\end{align}
while for $r=0$ one has
\begin{equation}
 \psi(0,t) = \frac{1}{L} e^{-i \epsilon_0 t} + \frac{1}{L} \sum^{N}_{k=1} 2^{k-1} \ e^{- i \epsilon_kt}.\label{eq:psi_in_the_first_site}   
\end{equation}

\begin{figure*}
    \centering
    \includegraphics[width=\linewidth]{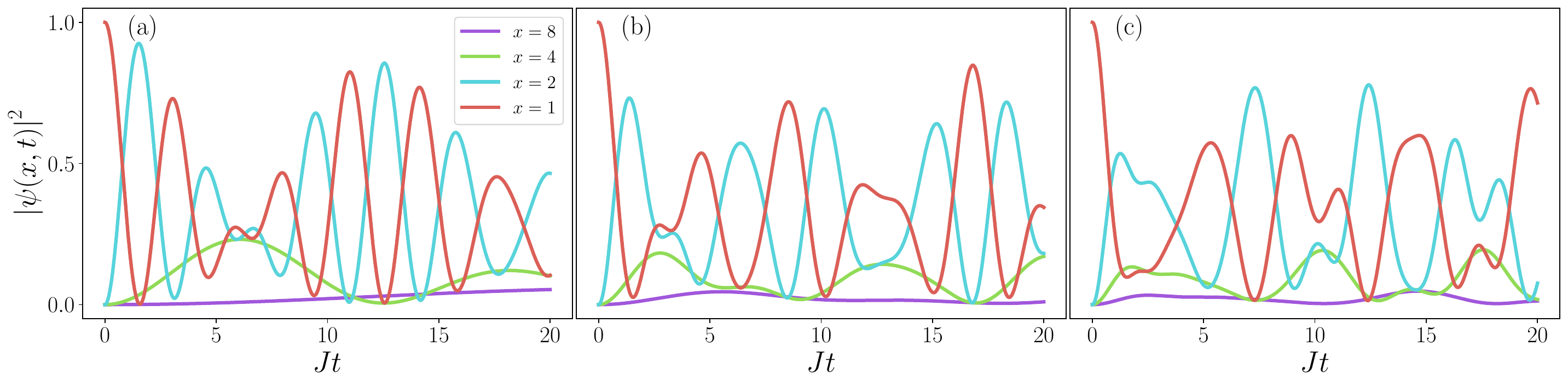}
    \caption{Analytical prediction for the single-particle  probability distribution, as a function of time $t$ and the site $x$, given by Eqs. \eqref{eq:An_r0} and \eqref{eq:recursive}. We show (a) $\sigma=2$; (b) $\sigma =1$; (c) $\sigma =0.5$}
    \label{fig:figure1}
\end{figure*}
\subsection{Scaling limit with $r$ fixed and $N\rightarrow +\infty$}

So far, we have derived an expression for the evolution of the state localized at the first position (see Eq. \eqref{eq:psi_generic_site} and \eqref{eq:psi_in_the_first_site}), which is exact for any length $L=2^N$ of the chain, as long as $h/J \rightarrow +\infty$. Now, we want to understand what happens in the thermodynamic limit 
$N \rightarrow \infty$.
It should be clear from the explicit expression of the sum appearing in Eq. \eqref{eq:psi_in_the_first_site} that the dominant contributions come from the large values of $k$, namely the high-energy (single-particle) modes. For this reason, it is rather natural to change variable
$k \rightarrow N-k$, so that the first terms of the sum become the relevant ones and one can thus approximate the sum as series in the large $N$ limit. Under this change of variable, the single-particle spectrum can be parameterized for large $N$ (up to an irrelevant additive constant) as
\be
\varepsilon_k = \tilde J_{\sigma} 2^{- \sigma k} + \text{const.},
\quad \tilde J_\sigma \equiv J\frac{2^{\sigma+1}-1}{2^{\sigma} - 1}.
\label{eq:spectrum}
\ee
We remark explicitly that $\varepsilon_k = \epsilon_{N-k}$, and we use a distinct symbol to avoid confusion. 
We now keep $r$ fixed, send $N\rightarrow +\infty$, and analyse how $\psi(r,t)$ behaves in time as a function of $Jt$. We first consider $r=0$, for which we have
\be
\psi(0,t) = \frac{e^{-i\epsilon_0t}}{L}+\sum^{N-1}_{k=0}2^{-k-1}e^{-i\varepsilon_kt}.
\ee
The sum above converges exponentially fast as $N\rightarrow \infty$ to a finite value that we write as
\be
\psi(0,t) \simeq \sum^{\infty}_{k=0} 2^{-k-1} e^{-it\varepsilon_k} = \sum^{\infty}_{k=0} 2^{-k-1} e^{-i \tilde{J}_\sigma t 2^{- \sigma k}}.
\label{eq:An_r0}
\ee
Similarly, for $r\geq 1$ we get
\be
\begin{split}
\psi(r,t) = \sum^{\infty}_{k=r} 2^{-k-1} e^{-it\varepsilon_k} - 2^{-r} e^{-it\varepsilon_{r-1}} =\\
2^{-r} \l \psi(0,2^{-\sigma r} t) - e^{-i \tilde{J}_\sigma t 2^{-\sigma (r-1)}} \r.
\end{split}
\label{eq:recursive}
\ee
The above analytical expressions are 
one of the main finding of our work, 
which in turn results in the localization of the initial spin-flip 
around its original position.

We show this in Fig. \ref{fig:figure1} where we plot, for the representative values $\sigma=0.5,1,2$, the time-evolution of the absolute square of the single-particle wave-function in the thermodynamic limit, for few lattice positions.

We notice that the exponential convergence of the series defining $\psi(0,t)$ is a feature that does not rely on the explicit expression of the single-particle energies $\varepsilon_k$ but only on the shape of the single-particle eigenfunctions. The latter ones are distinct from plane waves, which typically occur in translational invariant systems, and they are a feature of hierarchical models (displaying the symmetries of a Bruhat-Tits tree). A striking consequence is that $|\psi(0,t)|^2$, which is the return probability, oscillates in a typical time scale $\sim J^{-1}$ independent of the system size: this mechanism is already a feature of a localized system.

If we keep track of the precise details of the interactions, using $\varepsilon_k$ in Eq. \eqref{eq:spectrum} as single-particle spectrum, the dynamics shows additionally scale invariance. More precisely, we have that for $r>0$
\be
\label{scaling}
\psi(r,t) = 2^{-r}F\l t 2^{-\sigma r} \r,
\ee
with $F(t) = \psi(0,t) - e^{-i \tilde{J}_\sigma 2^{\sigma} t}$
being an universal function which depends on $\sigma$ and $Jt$ only. 
The scaling in Eq. \eqref{scaling} can be better understood if the hierarchical distance $r$ is compared to the euclidean one. Restricting the analysis to a position $x$ on the chain which is an integer power of $2$, say $x = 2^r$, we have that the hierarchical distance between $x$ and the first site is exactly $r$. This means that we can write Eq. \eqref{scaling} as
\be
\psi(x,t) \sim \frac{1}{x} F\l t x^{-\sigma} \r.\label{eq:universal_scaling_psiofxandt}
\ee
The scaling above makes transparent the presence of a dynamical exponent $z \equiv \sigma$, which relates time and space as $t\sim x^z$.
Finally, it is worth mentioning that the value of $z$ we observe is eventually related to the paramagnetic phase of the model (we are assuming $h\gg J$) and it differs explicitly from the dynamical exponent at the critical point (investigated in Ref. \cite{Monthus_2015}).

\subsection{Time averages and upper bound}

In the previous section, we did not use the specific details of the universal function $F$ appearing in Eq. \eqref{scaling}, besides that it is limited and it oscillates in time with a typical time scale $\sim J^{-1}$. This section aims to investigate exactly some features related to the time averages and the upper bounds of the probability distribution, to better quantify the localization properties of our dynamical protocol.

For any site $j$ of the lattice at hierarhical distance $r=r(1,j)$ from the first one, the probability that the particle is found at position $j$ at time $t$ is $| \psi(r,t) |^2$.
A straightforward computation shows that for $r\geq 1$ the number of lattice sites at hierarchical distance $r$ from the first site is exactly $2^{r-1}$. This means that the probability of finding the defect in a generic site at distance $r$ is given by 
\be 
 P(r,t) \equiv  \begin{cases}2^{r-1} |\psi(r,t)|^2 \quad &r\geq 1, \\ \ \\ |\psi(0,t)|^2 \quad &r=0.\end{cases}
 \label{eq:Prt_def}
\ee

Let us now consider the long-time average of the probability $P(r,t)$, defined as
\begin{equation}
    \langle P(r,t) \rangle \equiv \underset{T\rightarrow \infty}{\lim} \frac{1}{T} \int^{T}_0 P(r,t) dt.
    \label{P_time_av}
\end{equation}
We will show that for any given $r$, and in particular for $r=0$, this long-time average is finite in the thermodynamic limit $L\rightarrow \infty$. Let us first compute $ \langle P(0,t) \rangle $, for which we have
\be
    \langle P(0,t) \rangle =
    \frac{1}{4} \sum_{k,k'} 2^{-k-k'} \langle e^{-i \tilde{J}_\sigma t (2^{-k \sigma}-2^{-k' \sigma})} \rangle.
\ee
Since the single particle energies $\varepsilon_k$ differ for different $k$, the coherence terms vanish in the long-time average and one gets $\langle e^{-i \tilde{J}_\sigma t (2^{-k \sigma}-2^{-k' \sigma})} \rangle = \delta_{k,k'}$. This results in
\begin{equation}
    \langle P(0,t) \rangle = \frac{1}{4} \sum_{k=0}^{\infty} 2^{-2k} = \frac{1}{3}.
    \label{eq:time_av0}
\end{equation}
A similar calculation shows
\be
    \langle P(r,t) \rangle = \frac{2^{1-r}}{3}, \quad r \ge 1
    \label{eq:time_av}
\ee
As a consistency check, one can easily verify $\sum^{\infty}_{r=0} \langle P(r,t) \rangle = 1$.
From the results in Eqs.~\eqref{eq:time_av0} and \eqref{eq:time_av}, we learn that the particle can be found in average with probability $1/3$ both in the first and the second site (represented by $r=0$ and $r=1$ respectively), which is also the average probability of finding it at hierarchical distance $r\geq 2$. More in general, the
probability of finding the particle at hierarchical distance $r$ greater than $R$ is
\be
\la P(r>R,t) \ra\equiv  \sum^{\infty}_{r=R+1} \la P(r,t) \ra = \frac{2^{1-R}}{3},
\ee
which is exactly the average probability that it is at distance $r=R$ (see \eqref{eq:time_av}). We comment explicitly that if the time $T$, appearing in the definition of time-average in Eq.~\eqref{P_time_av}, is kept finite, then the agreement with our predictions (Eqs. \eqref{eq:time_av0} and \eqref{eq:time_av}) is expected to hold only up to a hierarchical distance $r$ such that $2^{r\sigma} \ll JT$.
As a consequence, the more the position is distant from the first site, the more the relaxation (on average) is slow. This observation explains why the long-time averages do not depend explicitly on the parameter $\sigma$, which may look like a counter-intuitive property at first sight.

Beyond the probability averages, we also provide some upper bounds on $P(r,t)$ valid for any $t$. Indeed, for $r>0$ we have get 
$|\psi(r,t)| \leq 2^{-r} 
\left( 1 + | \psi(0,t)| \right) \leq 2^{1-r},
$
where the triangular inequality has been used, together with $| \psi(0,t)|\leq 1$. Finally, one can show the probability of being at a hierarchical distance $r>R$ can be bounded by
\be
P(r>R,t)\leq 2^{1-R}.
\ee
As a matter of fact, the local excitation is localized around the first sites, not only on average but also for any time $t$ no matter how large it is.

\begin{figure}
    \centering
    \includegraphics[width=.9\linewidth]{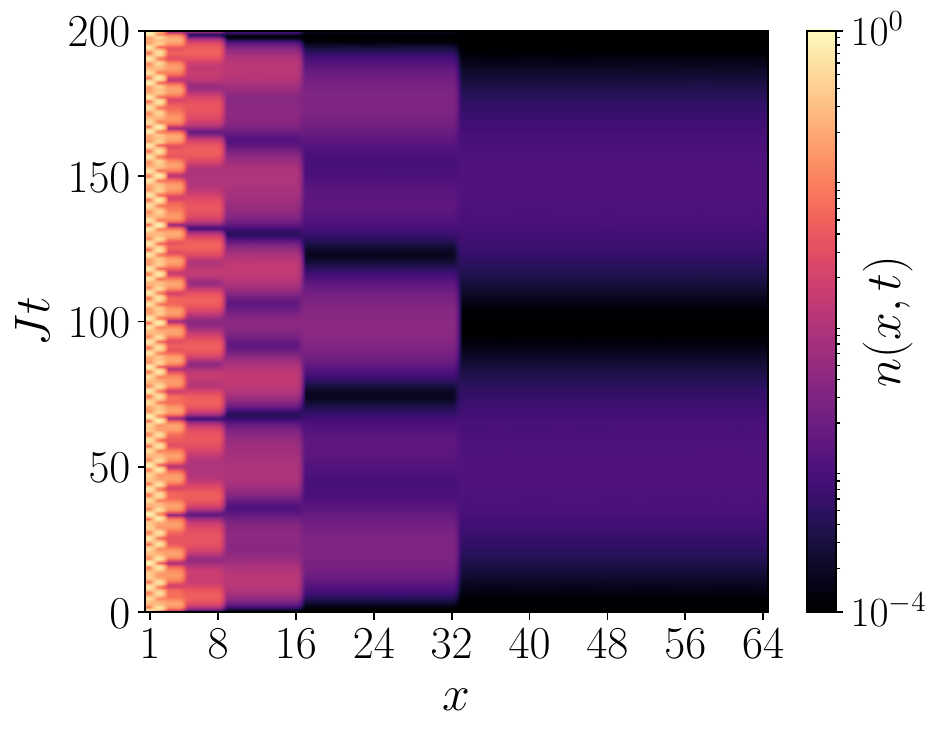}
    \caption{TDVP simulations of the average number of particles at position $x$ and time $t$. A chain of length $L=2^6=64$ has been chosen, and the parameters are $\sigma = 1$ and $h=40J$.}
    \label{fig:figure3}
\end{figure}

\begin{figure*}
    \centering
    \includegraphics[width=.49\linewidth]{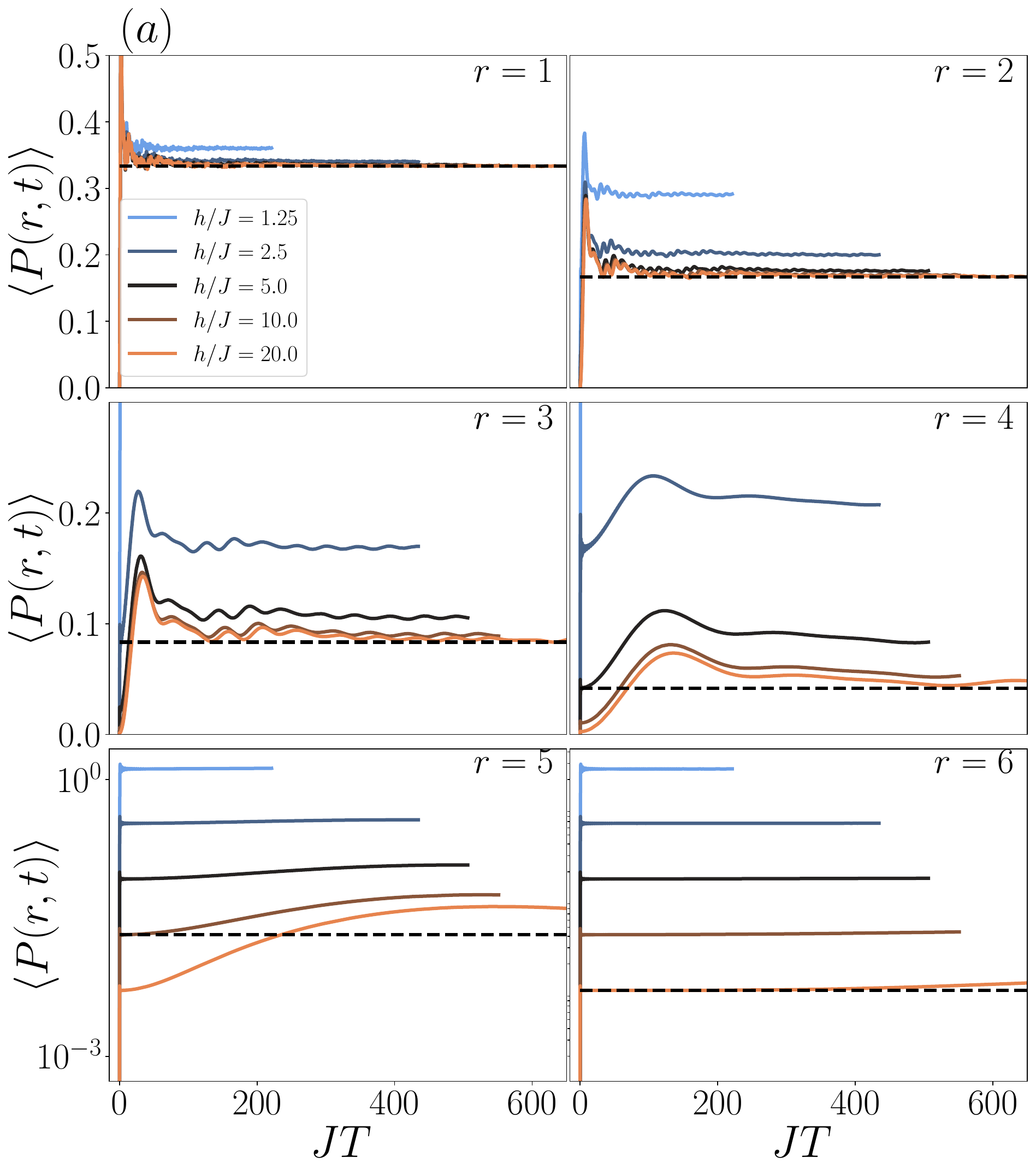}
    \includegraphics[width=.49\linewidth]{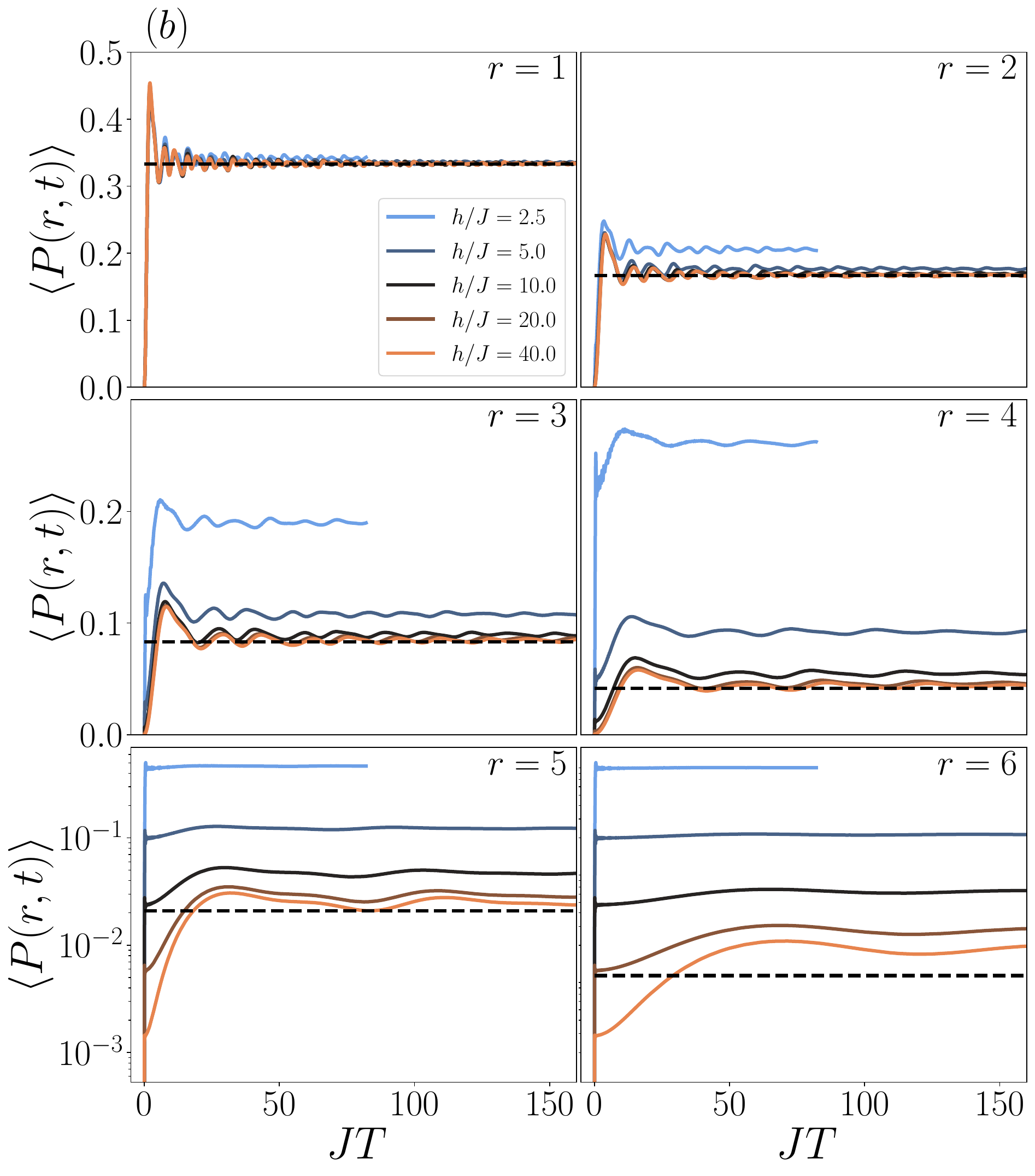}
    \caption{The average number of particles at hierarchical distance $r$ from the first site, averaged in a time window $t \in [0,T]$. Distinct values of $r$ are shown for a chain of length $L=64$. The values of $\sigma$ are (a) $\sigma = 2$; (b) $\sigma = 1$. The TDVP results (full lines) approach the analytical prediction for large $h/J$ and small $r$. Notice how the scale of the plots is linear for $r\leq 4$ while it is logarithmic for $r>4$, in order to enhance the small values of $\left<P(r,t)\right>$.
    }

    \label{fig:figure4}
\end{figure*}

\begin{figure}
    \centering
    \includegraphics[width=\linewidth]{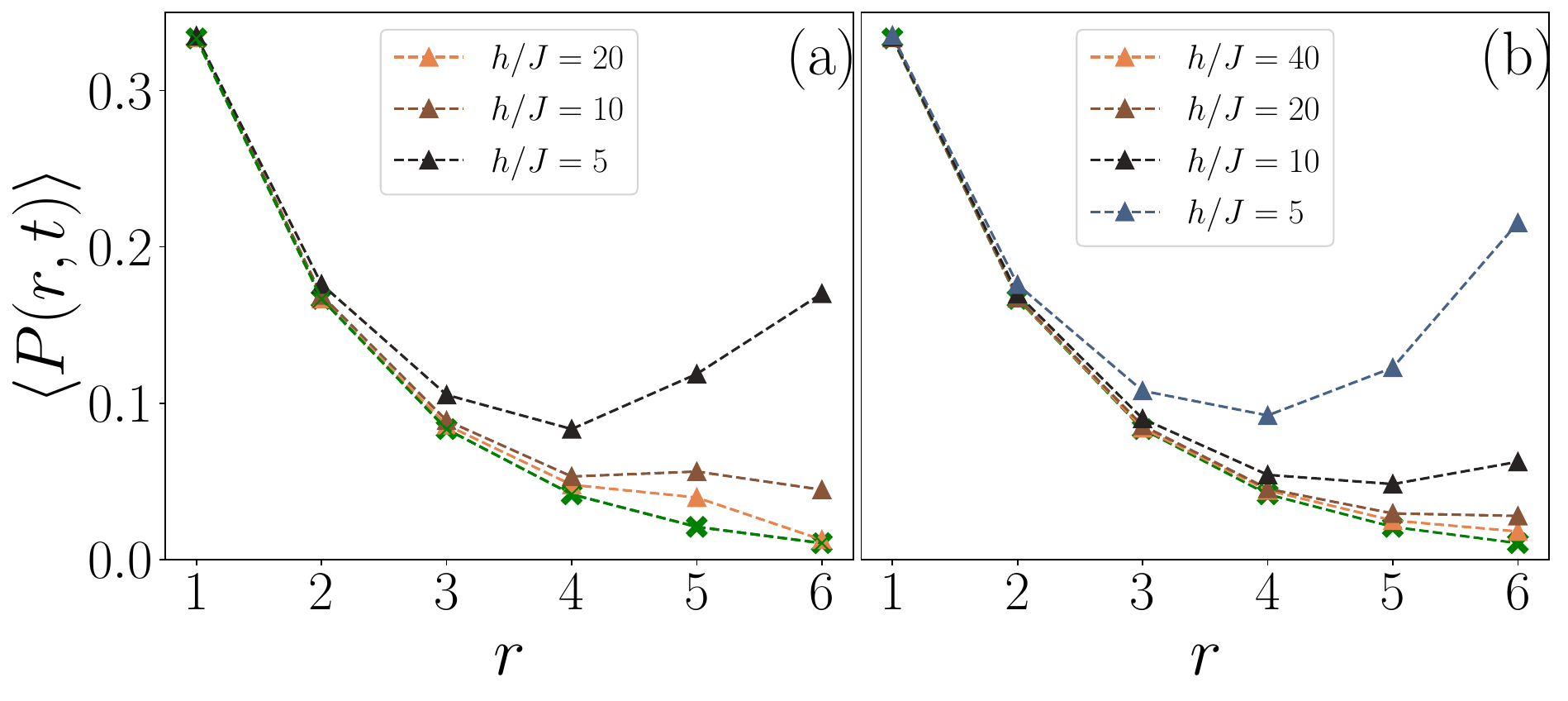}
    \caption{\commentas{As in figure \ref{fig:figure4} for large $JT$. The green line is the theoretical prediction obtained in the single particle picture. The number of defects increases for decreasing the value of $h$, thus showing deviations from the single particle picture.(a) $\sigma = 2$; (b) $\sigma = 1$.}}
    \label{fig:figure4_new}
\end{figure}


\begin{figure}
    \centering
    \includegraphics[width=.85\linewidth]{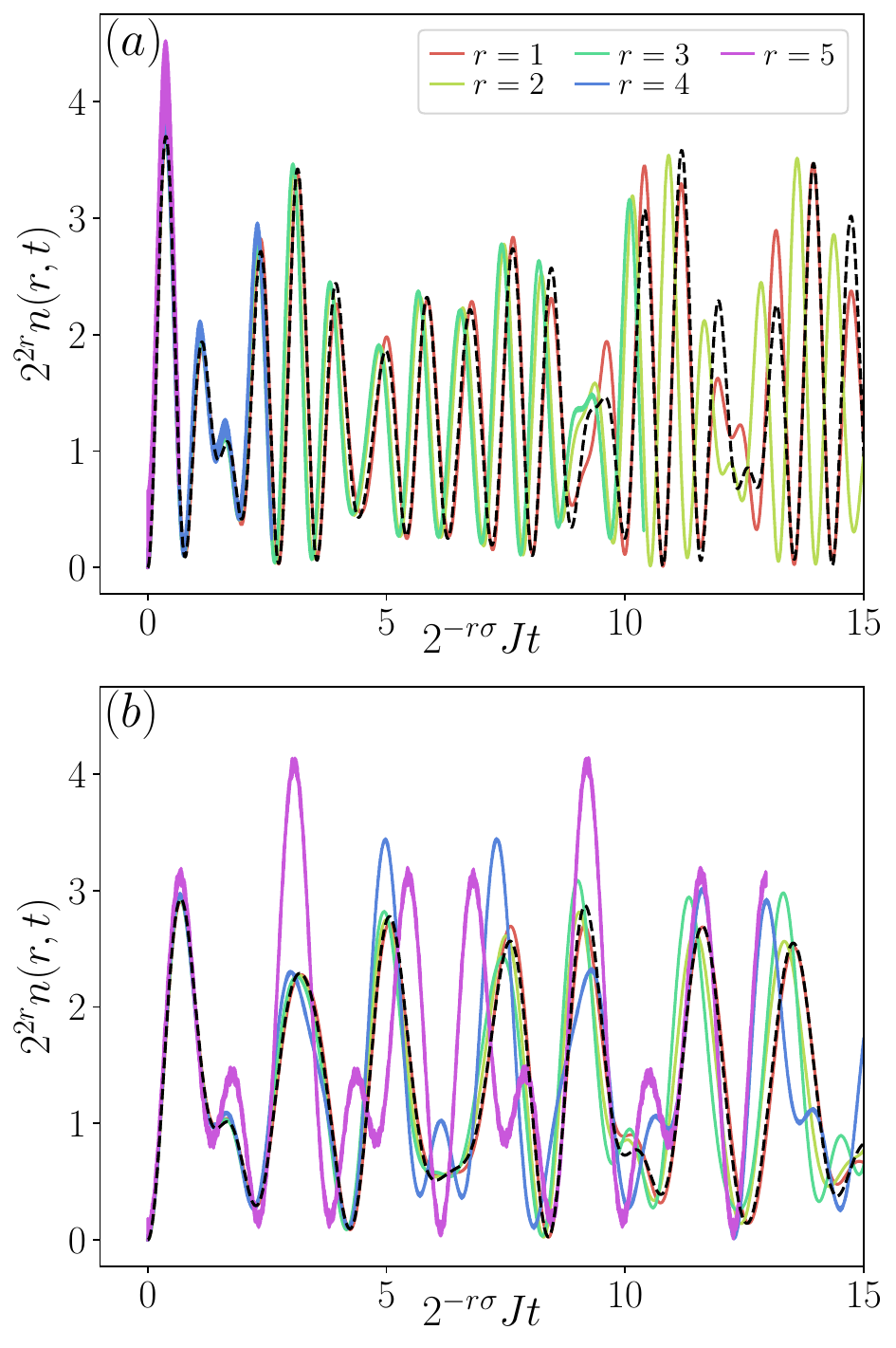}
    \caption{
    The figure shows the behaviour of $2^{2r}n(r,t)$, which is expected to be a universal function for large $h/J$, large $N$ and fixed $r$. The dashed line marks the theoretical prediction of the universal function, while the continuous line refer the numerical data for distinct values of $r$. We consider the parameters (a) $\sigma = 2$, $h=20J$; (b) $\sigma = 1$, $h=40J$. A general good agreement is found for small $r$, while bigger deviations are found at $r>4$, probably due mostly to finite size effects.}
    \label{fig:figure6}
\end{figure}

\subsection{Singular limit $\sigma \rightarrow 0$}
The predictions provided so far refer to any finite value of $\sigma>0$. Here we investigate the limit $\sigma \rightarrow 0$, which has to be handled carefully due to divergence at the level of the single-particle energies $\varepsilon_k$ in Eq.~\eqref{eq:spectrum}. Despite this divergence, we show that $|\psi(r,t)|^2$ has a well defined limit for $\sigma\rightarrow 0$ and $Jt$ fixed.

By expanding $\varepsilon_k$ for small $\sigma$, we get
\begin{equation}
    \varepsilon_k = J \left( \frac{1}{\sigma \ln 2} + \frac{3}{2} - k  \right) + O(\sigma),
\end{equation}
where the diverging term for $\sigma \rightarrow 0$ enters only as an additive constant, and it contributes as an irrelevant global phase on the wave-function $\psi(r,t)$. By getting rid of it, we find 
\be
\psi(0,t)|_{\sigma=0} = 
\frac{1}{2 -  e^{i J t} },
\ee
while for $r\geq1$
\be
\psi(r,t)|_{\sigma=0}
= 2^{1-r} e^{iJrt} \frac{1- e^{-iJt}}{2-e^{iJt}}.
\ee
Similarly, the probability reads 
\begin{equation}
    P(r,t)|_{\sigma=0} = \begin{cases}\displaystyle\frac{1}{5-4 \cos J t} \quad &r=0,\\ \ \\
    2^{-r}   \displaystyle\frac{4-4 \cos J t
}{5-4 \cos J t} \quad &r\geq 1.
    \end{cases}
    \label{eq:Psigma0}
\end{equation}
The latter results show that in this limit the evolution of $P(r,t)$ is periodic with a period $ 2 \pi/J$, which does not depend on $r$. This is compatible with the rough estimation of the recurrence time
$\sim J^{-1} 2^{r\sigma}$ coming from scaling arguments.

We finally emphasize that the finite value we get relies on the fact that $J$ has been kept fixed in the thermodynamic limit $N\rightarrow \infty$ and $\sigma \rightarrow 0$. As a matter of fact, it turns out that the critical value of $h/J$, above which the model is in the paramagnetic phase, goes to infinity as $\sigma \rightarrow 0$ (see \cite{Monthus_2015}). This means, in practice, that for small $\sigma$ we need huge values of the magnetic field in order to observe the single-particle regime we have described so far. One could in principle overcome this issue by a redefinition $J\rightarrow J\sigma$ (Kac's rescaling). However, with the latter prescription the typical relaxation time would be $\sim \sigma^{-1}J^{-1}$,
thus diverging for $\sigma \rightarrow 0$. As a consequence, the system would be frozen forever in the initial state for $\sigma \rightarrow 0$. These observations motivate that, despite the finite value of $P(r,t)|_{\sigma=0}$ in E.~\eqref{eq:Psigma0}, the limit $\sigma \rightarrow 0$ remains somehow pathological.

\section{Numerical Results}\label{sec:Num}
In order to explore the effects of finite $h$ we used a suitable matrix product state (MPS) representation of the many-body wave function, joined with the correspondent matrix product operator (MPO) representation of the Hamiltonian \cite{SCHOLLWOCK201196}. The non-equilibrium dynamics has been computed via the time-dependent variational principle (TDVP) algorithm \cite{PhysRevLett.107.070601,PhysRevB.94.165116,PAECKEL2019167998}.

\commentas{In particular, we considered the two-site TDVP algorithm with $dt=0.01$, the long-range MPO is constructed through a finite-state machine \cite{MPO_Reference}. Then, we compressed the MPO's bond dimension with singular value decompositions \cite{compressionMPO}, thanks to the hierarchical structure of the interactions the resulting bond dimension after the compression of less than $10$, thus it is greatly reduced. In our simulation we used a MPS's max bond dimension $\chi=64$, which in the worst case produced a truncation error of order $10^{-6}$. }

Thanks to the symmetries of the model and the quasi-conserved charge $S^z_{(N,1)}$, the entanglement produced during our dynamical protocol grows quite slowly, and the MPS description allows to reach very large times. In addition, to make a connection with the full many-body dynamics at finite $h$, we notice that the expectation value of the local number of excitations can be related to the local magnetization as follows
\begin{equation}
    n(x,t) =  \frac{1}{2}\left(1-\ev{\commentas{\sigma^z(x)}}{\Psi(t)}\right).
\end{equation}
while the many-body equivalent of the definition of $P(r,t)$ is
\begin{equation}
 P(r,t) = \begin{cases}2^{r-1} n(r,t) \quad &r\geq 1, \\ \ \\ n(0,t) \quad &r=0.\end{cases}
\end{equation}
As $h \gg J$, the single-particle picture is justified and we have $n(x,t) \approx |\psi(x,t)|^2$. This is the case of Fig. \ref{fig:figure3}, in which we show the evolution of $n(x,t)$ for $h=40J$ and $\sigma=1$.  In agreement with Eq.~\eqref{eq:universal_scaling_psiofxandt}, the dynamics is localized in the first few sites and the defect is unable to spread along the tree. Moreover, since $h$ is pretty large $S^z_{(N,1)}$ is well conserved. The sites which are further from the initial position of the defect, along the branches of the binary tree, show exponentially slow dynamics and suppressed wave-function modulus. 

In Fig. \ref{fig:figure4}  we show the evolution of the time-averaged probability $\left<P(r,t)\right>$ as function of $JT$, for some values of $h$ and $\sigma=1, 2$. The dashed horizontal lines marks the theoretical predictions for strong field given in Eq. \eqref{eq:time_av}. \commentas{In addition to this, in Fig. \ref{fig:figure4_new} we plot $\left<P(r,t)\right>$ as a function of $r$ for some values of $h$, for large enough values of $JT$ such that the average number of particle is settled to a stationary value.} As expected, for decreasing values of $h$, the number of defects along the chain is no more a conserved quantity, i.e.
\be
\sum^L_{x=1} \frac{1}{2}\left(1-\ev{\commentas{\sigma^z(x)}}{\Psi(t)}\right) \neq 1,
\ee
thus pair production becomes important, and the full many-body state is starting to leave the single spin-flip manifold. We note that for small values of $r<3$ the prediction holds also for values of $h$ not that far from the critical value $h_c$, i.e. $h_c(\sigma=2)\approx0.52J$ and $h_c(\sigma=1)\approx1.28J$\,\cite{Monthus_2015}. On the other hand, at large distances, and for small values of the transverse field $h$, the agreement with the single-particle approximation is getting worst since the pair production becomes the relevant mechanism in generating defects. 

Finally, in Fig. \ref{fig:figure6}, we inspect how the universality relation in Eq.~\eqref{scaling} is getting violated by varying the magnetic field $h$ and the hierarchical distance $r$.
Again, we find a good agreement with the single-particle prediction for small values of $r$ (a part from small deviations due to finite-size effects) up to larger re-scaled times $2^{-r\sigma} J t$, as far as we keep $h$ sufficiently large.

\section{Conclusions and outlooks}\label{sec:Con}

In this work, we analyzed a localization phenomenon in a hierarchical long-range model. We were able to construct analytically the single-particle wave function, describing the evolution of a defect in the deep paramagnetic phase, and we provided some bounds, together with the long-time averages, to give a quantitative description of the localization mechanism. Moreover, the universal scaling property in Eq.~\eqref{eq:universal_scaling_psiofxandt} has been found, which has been traced back to the self-similarity of the hierarchical tree structure.

This work paves the way to further possible investigations. There are still some open questions to be addressed indeed. In particular, from our derivation, it is not clear whether the localization is just a single-particle effect \cite{anderson1958} --- completely characterized by the eigenfunctions of the hopping matrix --- or a true many-body localization (MBL) effect \cite{mbl1,mbl2,mbl3}.

We believe that the symmetry of the hierarchical tree is a key ingredient for these phenomena, and a systematic investigation of its consequences at the level of the many-body spectrum deserves more attention. For instance, it is reasonable that the Hilbert space of this model is fragmented and a huge number of Krylov spaces are present \cite{Moudgalya_2021}. Moreover, another consequence of the tree symmetry could be the presence of local integral of motions (LIOM), which prevents the thermalization of the system, making the hierarchical model more similar to a disordered system rather than a translational invariant long-range one.

Another interesting direction could be the investigation of the dynamics after a global (or local) quench, near the critical point. That would be a way to probe the properties of the system in the middle of the spectrum, and not just in the single-particle low-energy band. However, this is a much more difficult protocol where we do not expect to find exact analytical predictions. One possibility to overcome this problem could be the application of SDRG techniques: however, it is still unclear how to adapt the formalism of \cite{Monthus_2016} besides the ground-state, to tackle systematically higher-energy states. Another possibility relies on the relation between this model and a $p$-adic field theory \cite{Heydeman:2016ldy,
Hung:2019zsk}, which could be a way to tackle directly the scaling limit at criticality.

Last but not least, it could be interesting to study a free fermionic counterpart of the hierarchical Ising model. We think that beyond the single-particle localization, which should show identical features for both models, also some many-body properties could be similar and eventually related to the same tree symmetry. The advantage of this approach is the presence of well-established free-fermionic techniques \cite{Eisler_2007} which could shed  some light on the quantitative characterization of the conjectured MBL phase of this model. Still, it is not clear, which techniques can be successfully applied to do the lack of translational symmetry.
In conclusion, we believe that a large class of hierarchical quantum models could actually show properties that are similar to the ones of disordered systems, and a better understanding of their common features may be a huge step in the solution of some open questions regarding MBL \cite{Abanin_2019}.

\section*{Acknowledgments}

All the authors are grateful to Silvia Pappalardi and Federico Balducci for valuable discussions. LC acknowledges support from ERC under Consolidator grant number 771536 (NEMO).

\section*{Data Availability}
The data that support the plots within this paper and other findings of this study are available from the authors upon request. The code is available at \cite{githubtn}.

\appendix
\section{Diagonalization of the hopping matrix}\label{app:Diagon}

In this appendix, we analyze the spectrum of the hopping matrix whose matrix element read
\be
    \mathbb{J}_{ij}  = \begin{cases} \displaystyle-\frac{J}{2^{(1+\sigma)(r(i,j)-1)}} \quad i\neq j,\\ \ \\ 0 \quad  \text{otherwise},
\end{cases}
\label{eq:hop_matrix}
\ee
for a finite value of the length $L=2^N$. Here we follow closely Ref.  \cite{Agliari-2016}, where a similar matrix was considered and the set of eigenvectors/eigenvalues was provided. A  remarkable consequence of the hierarchical structure of the hopping is that the set of eigenvectors does not depend explicitly on the values of the interaction terms at distinct levels, and it is completely fixed by symmetry arguments. This mechanism is somehow analogous to translational invariant systems, in which the single-particle spectrum is diagonalized in Fourier space.

 The lowest-energy eigenstate, identified by an index $k=0$, is given by the wave-function
 \be
\frac{1}{\sqrt{L}}\chi_{[1,L]}(x),
\ee
which represents single-particle state completely invariant under the symmetries of the tree. Similarly the wave-function
\be
\frac{1}{\sqrt{L}}(\chi_{[1,L/2]}(x)-\chi_{[L/2+1,L]}(x)),
\ee
associated to the index $k=1$, is an eigenstate which is invariant under the symmetries of the tree which do not mix the left/right half-chains and it is odd under the following permutation of sites
\be
1\leftrightarrow L+1,\dots, \frac{L}{2}\leftrightarrow L.
\ee
Beyond the previous eigenstates, there are multiplets of wave functions that generate degenerates eigenspaces of the hopping matrix. For instance, one can show that the states
\be
\begin{split}
&\sqrt{\frac{2}{L}}(\chi_{[1,L/4]}(x)-\chi_{[L/2+1,L/2]}(x)),\\
&\sqrt{\frac{2}{L}}(\chi_{[L/2+1,3L/4]}(x)-\chi_{[3L/4+1,L]}(x)),
\end{split}
\ee
associated to $k=2$, are exactly degenerate. Despite the fact that it is a matter of convention to choose them as a basis of the associated eigenspace or a linear combination of them, for practical applications it is convenient to choose them. Indeed, whenever a single-particle state is localized in the first half-chain its projection over $\chi_{[L/2+1,3L/4]}(x)-\chi_{[3L/4+1,L]}(x)$ vanishes identically, which means that the latter state does not participate explicitly to the dynamical evolution.
Similarly, one can show that a multiplet of the following four states ($k=3$)
\be
\begin{split}
&\sqrt{\frac{4}{L}}(\chi_{[1,L/8]}(x)-\chi_{[L/8+1,L/4]}(x)),\\
&\sqrt{\frac{4}{L}}(\chi_{[L/4+1,3L/8]}(x)-\chi_{[3L/8+1,L/2]}(x)),\\
&\sqrt{\frac{4}{L}}(\chi_{[L/2+1,5L/8]}(x)-\chi_{[5L/8+1,3L/4]}(x)),\\
&\sqrt{\frac{4}{L}}(\chi_{[3L/4+1,7L/8]}(x)-\chi_{[7L/8+1,L]}(x))
\end{split}
\ee
generates an eigenspace. The discussion can be straightforwardly generalized, showing that for $k\geq 1$ the $k$-th eigenspace has degeneracy
\be
2^{k-1}, \quad k=1,\dots,N.
\ee
Given the eigenstates of $-J_{ij}$, the computation of its eigenvalues $\epsilon_k$ appearing in \eqref{eq:eps_eig} is just a matter of algebra. Indeed, for $k=0$ one can show that the associated single particle-energy $\epsilon_0$ is just the interaction energy between the first site and the rest of the chain, namely
\be
\epsilon_0 = - \sum^{L}_{j=2} \frac{J}{2^{(1+\sigma)(r(1,j)-1)}},
\ee
which can be rewritten as a sum over sites at hierchical distance $r$ as follows
\be
\epsilon_0 = - \sum^{L}_{j=2}2^{r-1} \frac{J}{2^{(1+\sigma)(r-1)}} = -\frac{J}{1-2^{-\sigma}}\l 1-L^{-\sigma}\r.
\ee
Similarly, for $k=1$ one has to compute the interaction term between the first site and the other ones belonging to left chain, adding to this quantity the interactions with the sites of the right half chain with the opposite sign. In other words
\be
\begin{split}
\epsilon_1 = -\sum^{L/2}_{j=2}2^{r-1} \frac{J}{2^{(1+\sigma)(r-1)}} + \frac{L}{2}\frac{J}{2^{(1+\sigma)(N-1)}} =\\
-\frac{J}{1-2^{-\sigma}}(1-\frac{2^\sigma}{L^\sigma})+\frac{2^\sigma}{L^\sigma}.
\end{split}
\ee
Similar arguments can be applied for $k\geq 2$, and in the end the eigenvalues can be compactly written as
\begin{align}
\epsilon_k = -\frac{J}{1-2^{-\sigma}}\l 1 - \frac{2^{k\sigma}}{L^\sigma}\r + J \ \frac{2^{k\sigma}}{L^\sigma}\left(1-\delta_{k,0}\right),
\end{align}
for $k=1,\dots,N$. From the explicit expression of $\epsilon_k$ one notices that (at least for $\sigma \neq 0$), whenever $k\neq k'$
\be
\epsilon_k \neq \epsilon_{k'},
\ee
which means that distinct multiplets previously identified are not degenerate among each other. We emphasize that this is not a direct consequence of the tree structure of the hopping matrix, while it can be regarded as a generic property in absence of a fine tuning of the parameters $J_p$ appearing in \eqref{eq:hierarchical_hamiltonian}.\\
To conclude this appendix we observe that the decomposition of the delta function
\be
\delta_{x,1}
\ee
in terms of the eigenvectors (see \eqref{eq:delta_expansion}) contains exactly one basis element for each multiplet. This is the crucial property at the origin of this single-body localization. Indeed, as $k$ approaches to $N$, the eigenstates become more and more localized and their overlap with the delta function become non-vanishing in the thermodynamic limit
\be
N\rightarrow \infty, \quad N-k \text{ fixed}.
\ee
For this reason, the contribution of the localized eigenstates is the relevant one, while the one coming from the completely delocalized eigenfunctions (say $k=0,1$) is negligible.

\section{Entanglement Entropy}
In this appendix, we briefly discuss how the mechanism of localization affects the dynamics of the entanglement. To do that, we start from the prediction single-particle wave function in \eqref{eq:1part_wavef}, and we express the entanglement entropy in terms of it.

\begin{figure}
    \centering
    \includegraphics[width=\linewidth]{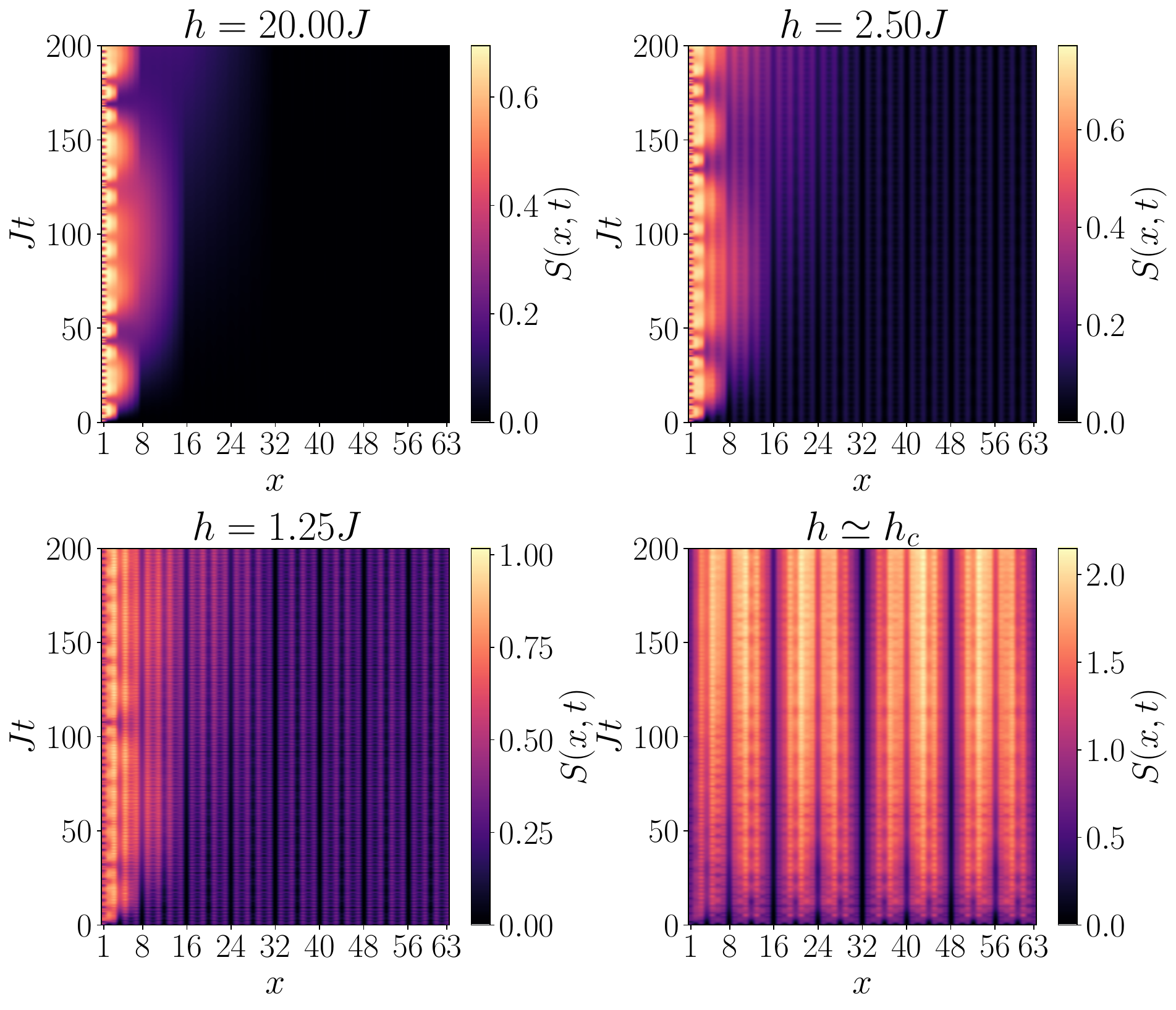}
    \centering
    \includegraphics[width=.9\linewidth]{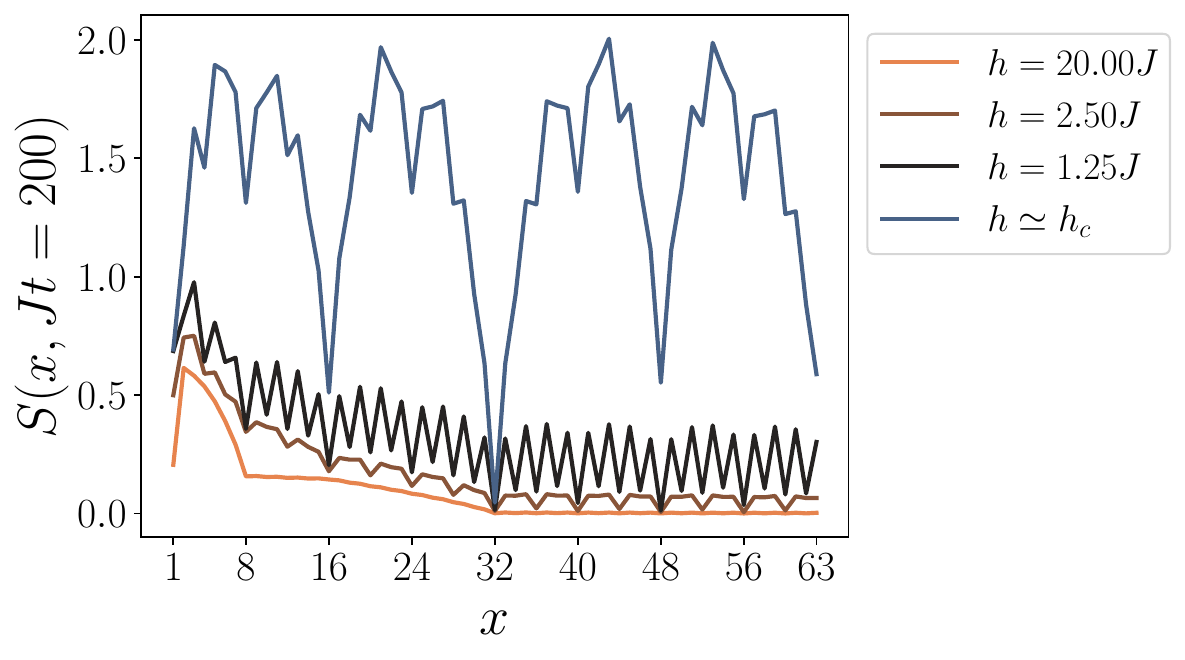}
    \caption{\commentas{Color plots:} Entanglement entropy $S(x,t)$ of the bipartition $[1,x]\cup [x+1,L]$, as function of $x$ and the time $t$. \commentas{Line plot: Entanglement entropy for fixed $Jt=200$.} The value $\sigma = 2$ and different values of $h/J$ have been plotted. The localization of the particle is manifest for large $h/J$, while the creation of pairs along all the chain is enhanced as $h/J$ approaches its critical value.}\label{fig:figure7}
\end{figure}

More precisely, let us consider a spacial subsystem $A$, made by a subset of the chain and a state $\ket{\Psi}$. The entanglement entropy is defined as
\be
S(A) = -\text{Tr}\l \rho_A \log \rho_A \r,
\ee
with $\rho_A \equiv \text{Tr}_{\bar{A}}(\ket{\Psi}\bra{\Psi})$ being the reduced density matrix of the state $\ket{\Psi}$. We now assume that the state $\ket{\Psi}$ is a superposition of states of the form
\be
\ket{\uparrow \dots \uparrow \downarrow \uparrow \dots \uparrow},
\ee
describing a single defect at the position $j=1,\dots,L$ of the chain. It is possible to show that in this case the entanglement entropy is precisely
\be
S(A) = -P_A \log P_A - (1-P_A)\log (1-P_A),
\ee
where $P_A$ is the probability to find the defect inside the region $A$. In other words, in terms of the single-particle wave function $\psi(x,t)$ we express $P_A$ as follows
\be
P_A \equiv \sum_{x \in A} |\psi(x,t)|^2.
\ee
Taking $A = [1,\dots,x]$ and using the scaling relation \eqref{eq:universal_scaling_psiofxandt}, we get that
\be
P_A = \sum^{x}_{x' =1} |\psi(x',t)|^2 \leq \sum^{x}_{x' =1} \frac{C}{{x'}^2},
\ee
for a certain constant $C$ which does not depend on the time. This upper bound for the probability  $P_A$ goes to zero as $x$ growth, and similarly for entanglement entropy. In particular a rough estimation gives the following scaling
\be
1-P_A \sim \frac{1}{x}, \quad S_A \sim \frac{\log x}{x},
\ee
for large value of $x$.

Here, we stress explicitly that the conclusion we made is based on the assumption that the one-particle effects are the only relevant ones and that the one-particle wave function shows localization properties at an arbitrary time.

In figure \ref{fig:figure7} we plot the evolution of the entanglement entropy obtained with TDVP for some values of the magnetic field $h$ and $\sigma = 2$. We observe that while for large values of $h$ the entanglement growth at large $x$ is suppressed, instead for smaller $h$ this is not the case. Indeed, for these values of $h$ entanglement is generated suddenly at arbitrary distances, which is due to pairs productions.

\end{document}